# Performance analysis of smart digital signage system based on software-defined IoT and invisible image sensor communication

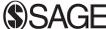

Mohammad Arif Hossain[1], Amirul Islam[1], Nam Tuan Le[1], Yong Tae Lee[2], Hyun Woo Lee[2] and Yeong Min Jang[1]

## Abstract
Everything in the world is being connected, and things are becoming interactive. The future of the interactive world depends on the future Internet of Things (IoT). Software-defined networking (SDN) technology, a new paradigm in the networking area, can be useful in creating an IoT because it can handle interactivity by controlling physical devices, transmission of data among them, and data acquisition. However, digital signage can be one of the promising technologies in this era of technology that is progressing toward the interactive world, connecting users to the IoT network through device-to-device communication technology. This article illustrates a novel prototype that is mainly focused on a smart digital signage system comprised of software-defined IoT (SD-IoT) and invisible image sensor communication technology. We have proposed an SDN scheme with a view to initiating its flexibility and compatibility for an IoT network-based smart digital signage system. The idea of invisible communication can make the users of the technology trendier to it, and the usage of unused resources such as images and videos can be ensured. In addition, this communication has paved the way for interactivity between the user and digital signage, where the digital signage and the camera of a smartphone can be operated as a transmitter and a receiver, respectively. The proposed scheme might be applicable to real-world applications because SDN has the flexibility to adapt with the alteration of network status without any hardware modifications while displays and smartphones are available everywhere. A performance analysis of this system showed the advantages of an SD-IoT network over an Internet protocol-based IoT network considering a queuing analysis for a dynamic link allocation process in the case of user access to the IoT network.

## Introduction

The Internet of Things (IoT) has become a buzzword and one of the dominant technologies in the globalized business market. It has brought radical changes in our everyday lives. It is assumed that the day is not very far away when devices can control themselves by making decisions and can even create interactivity among

[1]Department of Electronics Engineering, Kookmin University, Seoul, Korea
[2]Electronics and Telecommunications Research Institute (ETRI), Korea

**Corresponding author:**
Mohammad Arif Hossain and Yeong Min Jang, Department of Electronics Engineering, Kookmin University, Seoul, Korea.
Email: dihan.kuet@gmail.com; yjang@kookmin.ac.kr

devices through communication by identifying each other.[1] The interactivity has a high correlation with IoT as it is a concept of connecting things with anything, connecting a network to any network, and assigning an infrastructure to any infrastructure. IoT is a magnification of device-to-device (D2D) and machine-to-machine (M2M) technologies.[2] Furthermore, IoT has evolved as a megatrend in the automation industry, traffic management, industrial supervision, social security, disaster management, medical care, health care, and emergency services.[3–5] The world has observed revolutionary efforts in the sector of IoT from research institutes, the academic community, industrial divisions, and standard development organizations. The inspiration of IoT has been profoundly strengthened by all of these sectors, while the impact of IoT is very conspicuously exercised in advanced manufacturing initiatives.[6–8]

The term "smart world" is very closely interrelated to the advancement of IoT, as it develops the idea of establishing a connection among things, in any place, at any time. The world is becoming smart with the quick development of IoT. Our watches, bracelets, shoes, glasses, cell phones, and computers have become smart devices. Besides, transportation methods, such as a bus, car, motorbike, train, along with environments, such as home, office, fire service, medical, health-care center, and factory, have also turned to smart entities. Every aspect of our lives is becoming connected to the creative world, and the world is becoming more intelligent. This progress is greatly influenced by the rapid advancement of IoT technology.[9,10] Besides, the massive progression of IoT technology has inspired a smart ecological world, which will make the world more sustainable and more livable. Consequently, "IoT" is not bounded by the technological world, rather, it is anticipated to be considered as "green IoT."

The main idea of green technology has been developed based on the optimization of the usage of energy.[11] Green IoT has brought about the newly formed idea of green information and communication technology (ICT). The eventual consequences of green ICT principles are a cloud network along with software-defined networking (SDN) technology.[12] SDN is a new intelligent approach for the dynamic programmability of a network. Of late, it has emerged as a cost-effective and adaptive networking application. Furthermore, SDN technology has several advantages along with the benefit of making the IoT network a green IoT. The intelligence control mechanism of any system is highly associated with several routing and management protocols. Moreover, it is a very challenging task to change the protocol in every router or switch because the infrastructure of the Internet might be paralyzed. Moreover, numerous interfaces of any system of the Internet can make changing the routing protocols more convoluted and eventually make the supervision erroneous. SDN technology can bypass these problems to ensure the supportable evolution of a green IoT network and guarantee the proper management of the system.[13]

However, the interactivity among intelligent devices can make the world smarter when the devices can make their particular decisions and can be communicative with each other. The interactivity among devices can lead us into a new intelligent world where productivity can be increased to a great extent. The first and foremost condition for interactivity among devices is the establishment of communication between the devices. D2D communication is a technology in which the devices can communicate to set up interactivity between them. Moreover, the capacity problem of traditional cellular networks can be solved by adopting D2D communication. Furthermore, D2D communication can pave the way to use the out-of-band spectrum, which can improve the spectral problem along with the issues of capacity for a huge number of devices. The communication can be nonradioactive, which alleviates the detrimental effect of radio frequency (RF) wireless networks to humans and other living beings. Finally, the communication among devices can gain a higher energy efficiency, which is a vital concern for the implementation of green IoT.[14]

Inspired by the interactive green IoT, we have proposed a software-defined IoT (SD-IoT) architecture for a smart digital signage system where the interactivity between user and digital signage will be established by invisible image sensor communication (ISC), a mechanism for embedding data in the video image by watermarking that will be imperceptible to human eye but the camera sensor can detect the embedded data. The main idea of our proposed methodology is to develop a smart digital signage system using SDN where the digital signage can interact with users by smartphone camera and can connect the users to a cloud network using dynamic protocols for user access. In addition, the interactive environment for a smart digital signage system can be created simultaneously through the invisible ISC among the digital signage and the users during the ongoing advertisement. Consequently, this article has become a discussion of an SDN technology-based IoT paradigm for an interactive digital signage network, where the catalyst of the interactivity is the invisible ISC. The rest of the article is organized as follows: The current status and the problems of related works are described in section "Relevant works" while the architecture of the smart digital signage system is proposed and demonstrated in section "Proposed scheme." Section "Mathematical model analysis" includes the mathematical model for the queuing analysis. The performance of the proposed scheme is represented in section "Performance assessment," whereas section "Conclusion and future work" concludes our work with a discussion of future works.

## Relevant works

The research on interactive environments between humans and devices has become commonplace during this decade. The deployment of digital signage has come to be a very cardinal issue for creating interactivity that might bring a momentous advancement in the world of technology. It is predicted that the world will be embellished with digital signage in the near future and with other devices such as cell phones, tablets, smartphones, and wearable devices might interact with digital signage through communication. This possibility has introduced a new era of research into digital signage. However, several types of research have been published, and several are studying interactive digital signage. A cyber-physical broadcast/multicast (B/M) media system-based smart signage was proposed by She et al. The authors declared a protocol for multiple users with a view to acquiring content from digital signage using hand gestures.[15] This idea was asserted for multiple displays as well, considering the azimuth angle of a smartphone using the accelerometer and magnetic sensor of the smartphone. Interactions between smartphones and multiple signage displays that have been taken into account in this work are sensor-based techniques. A wireless router connects the user and the screen. Although this scheme includes the idea of simultaneous interaction and the mobility of the user, it makes no proposal for SD-IoT for interactivity between the digital signage and users.

A ubiquitous display application has been presented by Strohbach et al. for creating a platform for a retail environment.[16] The author discussed context-aware digital signage prototypes to establish a distributed context management framework (CMF) as well as an interactive display wall. The display wall was created as an eye-catcher while an radio frequency identification (RFID) tag was used under a CMF to display the selected content by the users. This prototype was based on the fusion technology of different sensors as well as a mounted camera on the display for user tracking. A targeted advertisement prototype has been developed based on the quick response (QR) code technology. Users can access another website using the QR code, which is mainly for retail applications. Although this method includes a proposal for interactivity between the user and display, it excludes the SD-IoT-based smart digital signage system. Furthermore, a visual-based matching and tracking algorithm was proposed by Hu et al. for creative interactivity between a screen and mobile terminals.[17] The camera of a cell phone has been used to create interactivity between a display and a human. A user can change the contents of the screen and drag the content if he or she wishes. Moreover, multiple users can edit or browse different items on the screen. This scheme is a fair illustration of interactive digital signage, but there is no scope for SDN or IoT concepts. A smart multimodal digital signage system has been introduced by Tung et al. for observing multi-party interaction.[18] The authors proposed an extensive display system along with a microphone, a video camera, and depth sensors to establish interactivity between different users using voices and gestures. Interactivity between devices is not included in this article.

However, some other research works have been conducted based on the IoT concept to create an interactive environment in a wireless sensor network (WSN). The interactive IoT network mainly depends on the system architecture and the protocol of the sensor network. Besides, the ISC is usually nonadaptive to a classical sensor network architecture and protocols. A multicast communication technology in a WSN has been suggested by Porambage et al. with a view toward establishing D2D and group communication for IoT applications. This work does not support an SDN-based approach to enable other protocols.[19] An illustration of M2M networks architecture has been shown for a low-power wide-area network by Xiong et al.[20] This work was proposed for WSN, which also has dependability on a fixed protocol and applicable for IoT concept. A cognitive-radio-network-based M2M communication method was proposed for IoT by Aijaz et al. in order to initiate an energy-efficient and intelligent network architecture.[21] Nevertheless, this effort has no compatibility to SDN or IoT, and the communication system has no relation to image sensor functionality.

## Proposed scheme

The criteria for any system to be smart contain supervision, adaptability, connectivity, pragmatism, and so on. Keeping these factors in mind, we have partitioned our proposed scheme into two main parts. The first part of our proposed methodology describes an SD-IoT technology for smart digital signage system that involves the SDN, a newly invented networking technology that can support dynamic and flexible networking by decoupling the control layer and data layer (i.e. physical layer) of a network. Furthermore, the SDN can give the opportunity of achieving a dynamic networking capability, improved management of quality of service (QoS), and digital content management to ensure better services for users in case of an increase in the number of users in the IoT network. The QoS management for the users such as acceptance of user request, minimization of delay time for linking the users with desired services, better utilization of total services is the cardinal purpose of our proposed SD-IoT architecture. However, an invisible communication technique has been suggested in the other part of our proposed scheme with a view to creating interaction

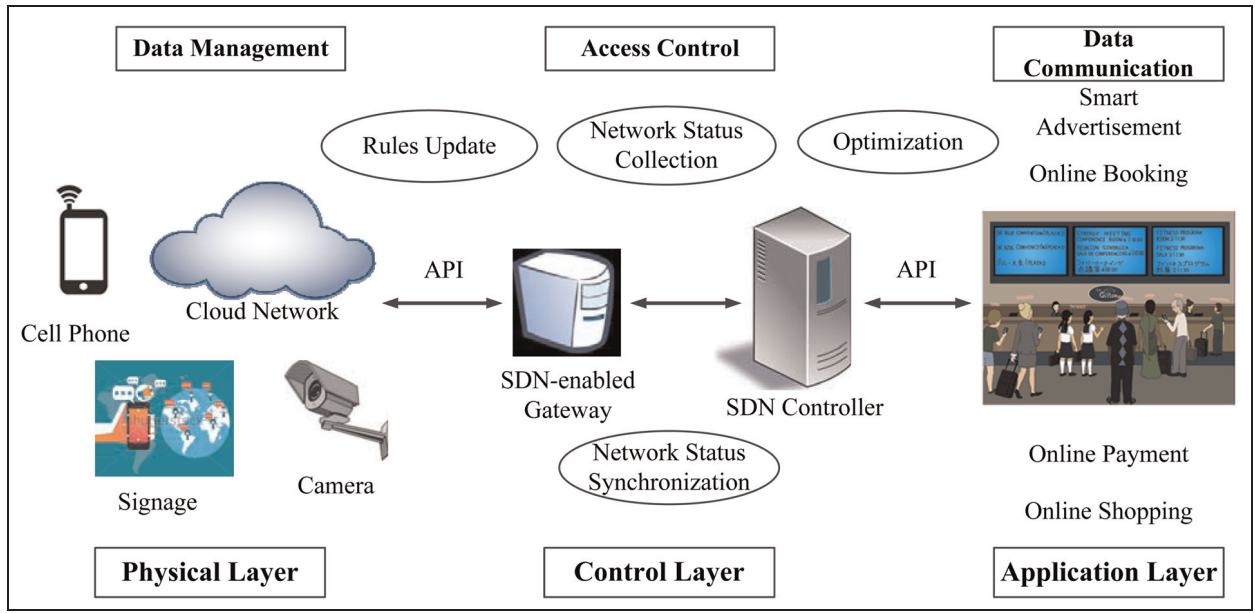

**Figure 1.** Proposed SD-IoT architecture for smart digital signage system. SD-IoT: software-defined Internet of Things.

among the users and the digital signages by employing the ISC technology for communication.

### Software-defined IoT

The vital proposal for our SD-IoT network is to create interactivity among the connected devices of a smart digital signage system in a cloud-based network as well as to control different layers of the network using SDN. The controlling of the network by SDN means decoupling the functionality of physical layer (e.g. devices) from the control layer. Besides, it supports automatic reconfiguration of any policy while the conventional networking, for instance, Internet protocol (IP)-based networking system, supports only static or manual configuration. It maintains interaction with different layer through an interface known as application programming interface (API) that provides SDN architecture the opportunity of directing the user applications, reducing the complexity of the network, and adjusting the bandwidth dynamically.

Our proposed SD-IoT architecture has been illustrated in Figure 1, which shows that the entire system can be divided into three primary layers: the physical layer, control layer, and application layer. The physical layer is a platform of physical devices that includes the nodes such as the transmitter, receiver, and management server. According to our proposed scheme, digital signage display is the transmitter while the receiver includes the image sensor used in the smartphone camera, handheld camera, and so on. The content management server consists of the cloud server, memory server, gateways, and so on. The function of this layer is to perform the transmission and the reception of data, that is, the data management. Consequently, this layer is also named as the data layer. However, the transmission of data between a digital signage (transmitter) and a smartphone (receiver) depends on the content management in the cloud server, as the data intended to transmit are inserted in the contents to display on the signage. The transmitted data might contain a link to a web address to provide services to the users such as online advertisements, online booking, online payment, and so on. The access of any user for any desired services can be controlled by API, an exclusive feature of SDN, effectively. One of the main advantages of using an API is to support multiple applications in physical devices. The features can reduce the complexity, maintenance costs, and can allow heterogeneity in the application. The API (e.g. OpenFlow, ForCES, OpFlex, etc.) protocol is mainly controlled by an SDN controller through an SDN-enabled gateway, which can be a virtual gateway. The core operation of the API in our proposed scheme is to allocate the desired bandwidth dynamically so that the users of the digital signages can be linked to their desired services in the network that results in an increase in the number of users.

The receiver (smartphone) can receive data from the digital signage using invisible ISC, a mechanism to transmit embedded data in the video image and to decode the transmitted data, from the displayed contents of the digital signage. The communication provides users with the links to the various web address for different services and connects the users in the cloud network. Nevertheless, the access of any user to any link on the Internet is not easy because there may be several users at the same time and users consuming the same advertisement from any digital signage will get the same data for linking to the web. Consequently, a mechanism is required to allow the users to access the link, which will result in easy content management for any advertisement producer.

Access to the link is controlled by the SDN controller of the control layer. The SDN controller can be considered as the heart of the SD-IoT architecture because it operates as the medium that connects the physical layer and the application layer. In our proposed scheme, the SDN controller is connected to the application layer and physical layer by different API interfaces. The interface between the application layer and the control layer is denoted as northbound interfaces while the southbound interface is considered as the interface between the physical layer and the control layer. The southbound interface (OpenFlow, ForCES, OpFlex, etc.) monitors the access of a user device in the physical layer, while the northbound interface allows the user in the application layer to provide the desired services. However, the OpenFlow protocol is the most modern southbound interface in the SDN. It can offer the SDN controller to maintain user access dynamically on the Internet. The primary functions of an SDN controller can be categorized into four types:

1. Logic design: The main application of an SDN controller is to transform the application specifications into packet forwarding rules. In addition, this function controls a protocol to establish communication between the application layer and control layer.
2. Rules update: The duty of an SDN controller is to generate packet forwarding rules in conjunction with properly describing the policies and installing the forwarding rules into compatible devices. However, the forwarding rules have to be updated with the policy changes according to the change of time. The SDN controller maintains consistency for packet forwarding rules after the updates are completed.
3. Network status collection: SDN controllers accumulate network status to provide a global view of the entire systems to the application layer. The network status includes duration time, packet number, data size, and bandwidth of flow. For instance, a helpful and commonly employed method for network statistics data collection is the Traffic Matrix,[22] which controls the volume of all traffic data that pass through sources and destinations in any network.
4. Network status synchronization: The prime goal of this process is to build a secure network using network status collected by individual controller because the unauthorized control of the centralized controller can degrade the performance of it. A general solution to overcome this is to maintain a consistent global view among all the controllers of the system.

The application layer is the layer in which the end-user applications can consume the data communications and network services. The physical structure of SDN can be shared by various applications at the same time, which can enhance the efficiency of the network. Our proposed architecture is capable of providing different kinds of services such as online booking, online payment, and online shopping. If an application needs on-demand connectivity as well as QoS guarantees (for instance, a bandwidth or channel), the SDN controller can estimate a qualified path through multiple layers and dynamically launch the connection by training the connection onto existing or newly established paths, as SDN controller has its global map information for multiple layers. The globally optimized path creation with QoS guarantees is not possible without the presence of an intelligent central controller. SDN can play a pivotal role here to fill the gap between design and implementation in multilayer optimization in optical networks. Thus, dynamic access in the cloud network is entirely feasible using SDN in the IoT network, which can ensure bandwidth utilization as well as capacity.

We have proposed a dynamic link allocation process for the SD-IoT architecture. The schematic diagram has been depicted in Figure 2. The mathematical model analysis for the system has been analyzed based on the proposed model in the next section (section "Mathematical model analysis"). There are $D$ numbers of digital signage displays belong to a cloud network in which every digital signage is connected to the network through the Internet connection (Wi-Fi connection) according to our proposed method. However, the advertisements displayed on the digital signage will carry information (link to some web address) to the users through the ISC technology. The connections of the users to the web addresses will be established by the cloud server too. There might be several accessed by the users in the same web address. Consequently, there need some reserved link channels for the users of each digital signage in the cloud server. Let us assume that the number of link channels, $U_d$ has been reserved for connecting the users of the display $d$ to the cloud. The number of reserved channels for each display is assumed to be equal, that is, $U_1 = U_2 = \ldots = U_d = \ldots = U_D$. Nevertheless, the number of users in each display may or may not be the same. Therefore, the number of users in each display is assumed as to be unequal. However, the SDN controller can estimate the network status and update the rules and can check the number of accessed user in the cloud server for each digital signage display. If there are more users in any display $d$ than other displays at any time, and there remain some unused channels belonging to other signage display in the cloud, the SDN controller can update the rules and allow the users to access the link channels by allocating the unused channels dynamically to the users waiting for the connection in web address.

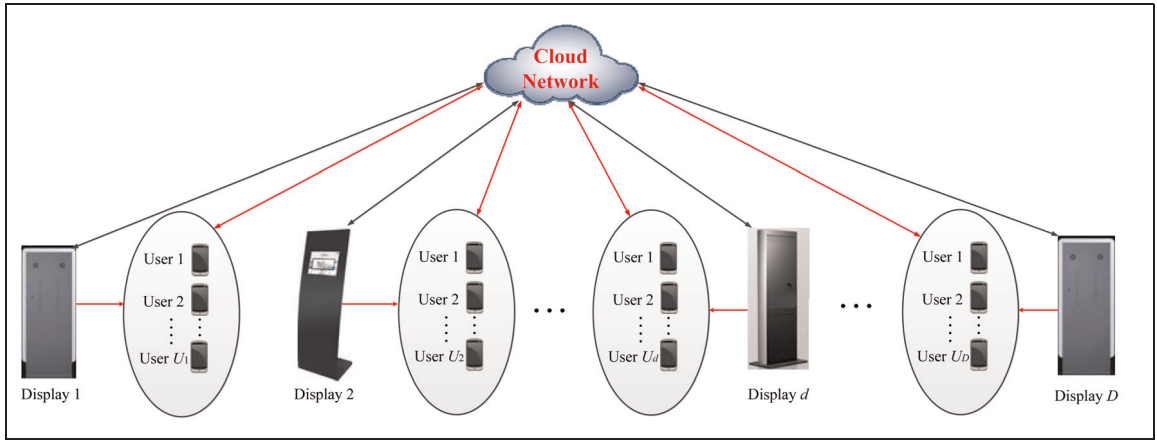

**Figure 2.** Scenario for dynamic link allocation process in smart digital signage system.

## Invisible ISC

Invisible ISC has a far-reaching effect in our proposed methodology, because the data (for instance, link address) is transmitted to the user using this technology. It has made our smart digital signage systems more appealing and effective. It is a novel form of communication for creating interactivity between a user and a digital signage and works as the catalyst in the smart digital signage system. One of the main reason of using invisible ISC technology for data communication is to use the existing digital signage infrastructure as a transmitter without bringing any hardware modification, because the data is embedded in the video image of the content that is displayed on a digital signage display. In addition, camera device (receiver) is also pervasive such as in smartphone camera that can be a good choice for creating interactivity between the user and digital signage.

However, the most challenging task for invisible communication technology is to create data that is hidden inside the image. Furthermore, there are possibilities of image distortion and noise insertion in the data, because the data is transmitted from digital medium to analog medium (i.e. the transmitter (digital signage display) is operated as the digital medium for hiding data, the channel medium is analog) and the data is received in the receiver (smartphone camera) also acts as the digital medium that interprets the conversion data from analog to digital form. Consequently, there are many opportunities for error during the period of communication. In addition, the geometric problems of the camera such as rotation, tilt, and mobility should be taken into consideration. Photometric distortion (e.g. distortion due to illumination) can be an issue in a lighting environment such as a large shopping mall or store.

Considering all of the complications mentioned above, we have proposed an invisible communication technology based on the intensity modulation of an image. Any slight change in the intensity of an image can be detected easily by camera while the human might be totally unaware of the changes. Besides, the change of intensity in the image is less complicated than the other methods that have been proposed previously for display-camera communication such as visual MIMO base methods.[23,24] In addition, the geometric effect which is one of the major problem for visual MIMO-based method can be eliminated effortlessly because the change of intensity of an image remains with change in the geometric position of a camera device. Furthermore, the photometric effect is negligible to a certain extent because the data is extracted by comparing two consecutive image frames of the transmitter and the illumination changes affects both image frames similarly. Therefore, the total difference in the intensity of the two consecutive image remains similar to the non-illuminated image.

Every pixel of an image has three color parts (red, green, and blue (RGB)) in proportion to the intensity value. We have proposed that an insignificant change in the intensity of any color of an image can be unrecognized by the human eye, while this minimal modification can be detected easily with a proper detection algorithm by the camera sensor in the smartphone. Moreover, we have proposed that any change brought in the intensity of the blue color part of an image remains more undetectable than the change in the intensity of other two colors (i.e. red and green) to the human eye. Accordingly, we have changed the intensity of blue part more than the intensity of other parts for the invisible communication.

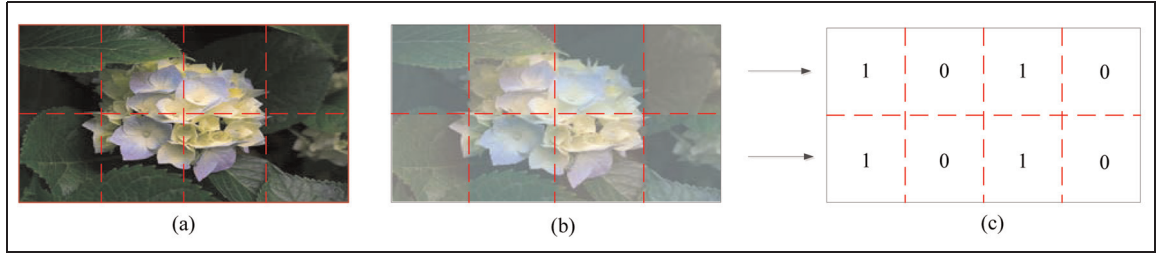

**Figure 3.** Segmented image to embed data by changing the intensity. (a) Image segmentation, (b) augmented camera view to realize the change in intensity, and (c) data acquisition.

We have partitioned an image into eight segments for communication purposes which can be compared to the first step of data embedding procedure. Each segment is a cluster of pixels, as illustrated in Figure 3(a), where the segmentation is highlighted using red dot line. The partitioning of the image segment has been performed in such a way that the number of pixels in each segment remains same or almost same. According to our scheme, if we need to send a message with bit "1," the pixel value of each pixel in a segment is increased or decreased. The intensity values of the pixels of a segment are kept unchanged which indicates the bit "0" if the information of the communication interprets such data. These steps are similar to data embedding procedures. Figure 3(b) shows a prototype to explain the change in the intensity of different pixel blocks with respect to the camera view and the change is unnoticeable to the human eye. The changes in the intensity have been converted into data that are represented as 1 or 0 according to the changes which is depicted in Figure 3(c). Besides, the arrow shows the direction of data counting in the figure. We have counted the data by going sequentially through each row, and row by row. It is obvious from Figure 3(c) that the data are "10101010." The technique of achieving output shown in Figure 3(c) can be termed as data acquisition procedure.

Our data decoding procedure is related to comparison of two consecutive images taken by a camera to observe the change in the intensity of the image. One image frame is considered as the reference image in which no changes are brought to the pixels of the image while the other image is the data embedded image in which the intensity of the intended pixels is changed. Thus, it becomes quite feasible to detect the change of intensity in the image. Although the mechanism of comparing reference image and embedded image can reduce the data rate by half, it is viable to provide services with low data rate such as address link of web, coupon, token, and so on. Because a smartphone camera has a shutter speed of 30 frames per second generally that interprets that it can capture 30 image frame in every second. Therefore, if we embed 8-bit data in each image frame, the smartphone can get 120 bits (i.e. 1/2 of (30 $\times$ 8), as other half of the images are used as reference image) data per second. Hence, it is quite reasonable to acquire a link address within few seconds. However, a prototype of our implemented invisible ISC technology has been shown in Figure 4, where the right display and left display are showing the original video image without data and the modified video image with embedded data, respectively. Nevertheless, the images in the left and right displays look very similar and show no distortion effects.

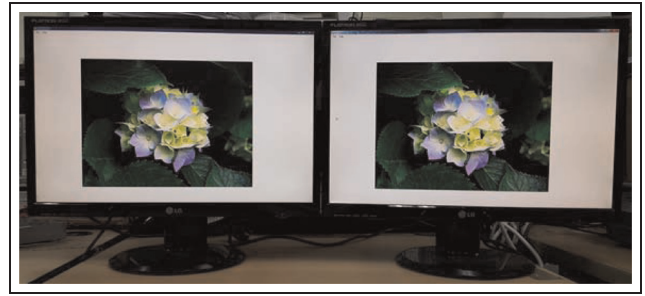

**Figure 4.** Demonstration of implemented invisible embedded data for smart digital signage system. (Right side display contains an original video image while the left one includes the embedded data inside the image of the video).

## Mathematical model analysis

Our proposed SD-IoT-based smart digital signage technology introduces interactivity between digital signage and users by getting a link ID (e.g. web address) through invisible ISC. The users of a digital signage can access the link by connecting to the cloud network. Our proposed system is capable of sending link information to many users simultaneously, as there can be numerous users in front of a digital signage at the same time. But the capacity of connecting the vast number of users to the link is inadequate, as the bandwidth of any network is limited. Nonetheless, our proposed SDN-based dynamic link channel allocation process is capable of connecting more users in the link channels than the conventional networking system. According to our proposed scheme, there will be some reserved channels for the users of every digital signage in a cloud network to access the web address. This circumstance generates a challenge to connect the intended users to the accessible link for a particular period. However, SDN has made it easy to allocate the channels to the users by initializing its protocol, which can be dynamic and more flexible with respect to the IP-based networking system.[25] With a view toward establishing an optimization process between a request from a user to access a link and the available unused link channels, we have proposed a dynamic link channel allocation process for the link-requesting users in the cloud network. The remaining link channels belong to any digital signage display may or may not be the same depending on the access rates of the users that can be changed over time to time. The number of unused channels for the users of any digital signage can be increased or decreased dynamically by borrowing or lending unused channels to the users of other digital signages using an SDN protocol, for instance, OpenFlow protocol. The process can be accomplished by creating a dynamic protocol in the SDN controller.

With a view to examining the impact of dynamic link channel allocation process for the users, a queuing analysis has been performed in this section by assuming $M/M/S$ queue model for simulation of the smart digital signage system. The queuing model is based on Poisson distribution process where there are an infinite number of users waiting in the queue to connect to link channels in the cloud server. Consequently, the request arrival rate becomes independent of the number of users, and it can be assumed to be the same in every state of the allocated channels. By contrast, the release or discharge rate of the users from the occupied link channels is related to the number of channel states in the queue. Our proposed dynamic link channel allocation process for the requesting users of a digital signage display will work, if there are available unused link channels reserved for the users of another digital signage display. The Markov models for our queuing analysis have been shown in Figures 5 and 6, which show the model before the allocation process of the channels to the queued users and the model after allocation process of the unused channels to the requested users of the queue, respectively.

Let us assume that the arrival rate of a request from the users belongs to a digital signage display, $d$ is denoted by $\lambda_d$, while the discharge rate of the occupied channels is indicated by $\mu_d$. The total number of digital signage displays occupied under the cloud network is $D$, and the reserved link channels for the users of the digital signage display, $d$ before the dynamic channel allocation process (i.e. the allocation of unoccupied channels to the requested users) is symbolized by $U_d$. Besides, the total number of channels after the allocation of the unused channels to the requested users is represented by $U_d'$, which interprets that $U_d'$ will be greater or lesser than $U_d$ depending on the borrowing or lending of channels, respectively. However, the probability of keeping a request in a queue from any user of the digital signage display, $d$, to connect in the reserved link channel, $m$ is $R_d(m)$, can be expressed by equation (1) when there are available reserved channels in the cloud for the digital signage. The queuing probability can be represented by equation (2) when there are no available channels unoccupied in the cloud to connect a user:

$$R_d(m) = \frac{\lambda_d^m}{\mu_d^m m!} R_d(0); \qquad m \leq U_d, \tag{1}$$

$$R_d(m) = \frac{\lambda_d^m}{\mu_d^m U_d! U_d^{m-U}} R_d(0); \qquad m > U_d, \tag{2}$$

where $R_d(0)$ represents the probability, $R_d(m)$ at the initial state which has been illustrated in equation (3):

$$R_d(0) = \left\{ \sum_{m=0}^{m=U_d-1} \frac{\lambda_d^m}{\mu_d^m m!} + \frac{U_d \lambda_d^{U_d}}{\mu_d^{U_d-1} U_d! (U_d \mu_d - \lambda_d)} \right\}^{-1}. \tag{3}$$

The probability of rejecting a request of a user in the queue belonging to signage display, $d$ is the summation of the queuing probability of the request when there are no available channels for the user, and it can be termed as $R_c(d)$, as shown in equation (4). Besides, the overall probability of rejecting a request of the entire system, $R_C$, can be expressed by equation (5).

$$R_c(d) = \sum_{m=U_d}^{m=\infty} R_d(m) = \frac{U_d \lambda_d^{U_d}}{\mu_d^{U_d-1} U_d! (U_d \mu_d - \lambda_d)} \left[ \sum_{m=0}^{m=U-1} \frac{\lambda_d^m}{\mu_d^m m!} + \frac{U_d \lambda_d^{U_d}}{\mu_d^{U_d-1} U_d! (U_d \mu_d - \lambda_d)} \right]^{-1}, \tag{4}$$

$$R_C = 1 - \frac{\sum_{d=1}^{d=D} \lambda_d [1 - R_c(d)]}{\sum_{d=1}^{d=D} U_d}. \tag{5}$$

The dynamic allocation process has created an option to utilize the available reserved link in an efficient way, which can ensure an escalation for the link utilization factor, $L_c$, a terminology to express the ratio of the used resources (channels) to the total available resources, has been presented in equation (6). The symbols used in the mentioned equations has been listed in Table 1 with their definition to make it easy for the reader. However, the scenario illustrated in Figure 2 inherits the system architecture for the queuing analysis process in the smart signage system. The figure will help the reader to understand the Markov models that have been presented in Figures 5 and 6.

Table 1. Denotation of the symbols.

| Attributes | Definition |
| --- | --- |
| $D$ | Total number of signage display connected in a cloud server |
| $D$ | Any signage display under the cloud network |
| $L_c$ | Link channel utilization factor |
| $M$ | Any link channel in the cloud |
| $R_d(m)$ | Probability of connecting any user belongs to signage display $d$ in the link channel $m$ |
| $R_c(d)$ | Rejection probability of a request from the user belongs to signage display $d$ in the link channel $m$ |
| $R_C$ | Overall rejection probability of the user in the cloud system |
| $Q_c$ | Link request rate in the cloud from the user belongs to each signage display |
| $Q_{c\_normalized}$ | Ratio of link request rate the user of each signage display to the overall rejection probability |
| $U_d$ | Number of reserved link channels in the cloud for the users of signage display $d$ before allocation process |
| $U_d'$ | Total link channels for the users belongs to signage display $d$ after allocation process |
| $\lambda_d$ | Request arrival rate in the cloud from the users belongs to display $d$ |
| $\mu_d$ | Release rate of the user belongs to display $d$ in the cloud |

$$L_c = 1 - \frac{[1 - R_C]\sum_{d=1}^{d=D}\lambda_d}{\sum_{d=1}^{d=D}U_d\mu_d}. \quad (6)$$

## Performance assessment

The performance evaluation section of our proposed scheme can be divided into two parts: simulation results for the queuing model analysis and the experimental results for the invisible ISC technology. On the one hand, our queuing model includes the performances of the request arrival rate, link access rejection probability, average delay, the average number of request in the queue, and, finally, the link utilization factor. On the other hand, the performance of the invisible ISC represents the prototype of our implementation by analyzing the similarity between the original image and the data-embedded image to keep the communication process invisible to the user. However, we have illustrated the performances of the queuing analysis in terms of link access rejection probability, link utilization factor, average delay, and an average number of requests in the queue with respect to the request arrival rate by making a comparison between an IP network and a SD network. These results clarify an overview of the advantages of an SD network over an IP network. In addition, an examination of the RGB histogram has been performed in order to compare the differences of the color intensities of the pixels of an original image and a data embedded image that have been used in the implementation of our invisible ISC.

The simulations performed in this section have been assisted by some assumptions. Firstly, the total number of digital signage displays under a cloud network has been assumed to be three for the simplicity of the simulation, while the total number of link channels that are reserved for the users of the every digital signage display has been considered to be equal by the number 10. Moreover, the average duration that a user spends at the link channels has been taken as 10 s. Furthermore, the simulation results for the queuing analysis have been accomplished by varying the request arrival rate from 0 to 10. The ratio of request arrivals rate for the three digital signage has been considered to be 5:3:2. Table 2 summarizes the parameters that have been assumed for the simulation purposes. However, one of the prime ideas of our proposed scheme is that the unused link channels of a display can be allocated to the users of digital signage that has more users (i.e. insufficient reserved channels with respect to users). This mechanism can result in a lower probability of link access rejection, while the utilization factor of the link channels can be augmented to a certain extent.

The link access rejection probability is the probability of rejecting a request from a user for web access as shown in Figure 7. This result has been analyzed by varying the request arrival rate. It is evident from the figure that a low request arrival rate has no effect in the rejection of request of the users in the cases of both an IP network and an SD network as there are available reserved channels to connect the users. However, when the request arrival rate increases, the situation changes and the request for the link channels surpasses the capability of the system. Nevertheless, the rules updating and network status collection characteristics of the SDN technology have bestowed the SD network with the dynamic link allocation capability. As a result, the SD network support more user than the IP network when the request arrival rate increases. Because of the dynamic allocation characteristics of SDN technology, the SD network is capable of allocating the unused channels of a digital signage to the requested users of another digital signage. Consequently, the SD network has the capability of achieving lower link rejection probability than the IP network because the IP network is not capable of collecting the network status and updating the rules from time to time.

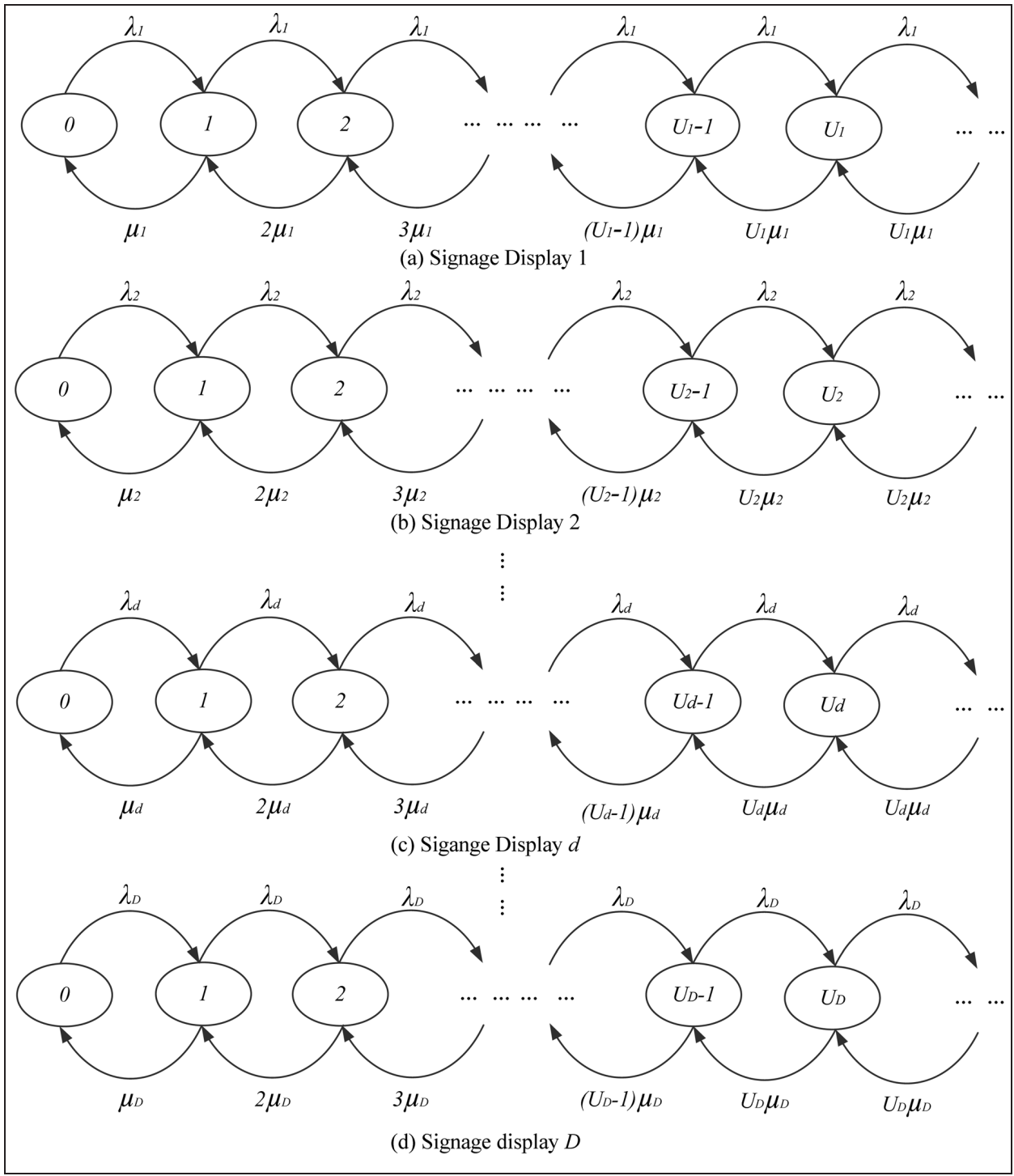

**Figure 5.** Pre-allocated Markov chain analysis for requesting user.

Additionally, the SD network has a greater link utilization factor than the IP network according to our proposed scheme. Figure 8 illustrates the comparison of SD and IP network with respect to link utilization factor. The link utilization factor can be defined as the percentage of link channels that are utilized link channels, and it interprets the usage of available link channels. However, the dynamic link allocation process for requesting users in the queue of an SD network increases the overall link utilization factor. Because there might be some unused channels that are reserved for the user of a digital signage but the SD network can collect the status the network and reallocate the unused channels to the queued users of other digital signage by updating its rule whereas the IP network possesses no characteristics of network status collection and rule updating mechanism. The Figure 8 shows that lower arrival rate of a request can increase the utilization of channels to a great extent as the lower request arrival rate specifies that some channels remain unused, and the SD network can allocate the remaining channels to the queued users. Consequently, the link utilization factor is intensified. Although a higher request arrival rate indicates less difference in link utilization between an IP network and an SD network, because the high request arrival rates represent that the reserved channels are occupied by the users. Thus, the reallocation process of the channels does not work.

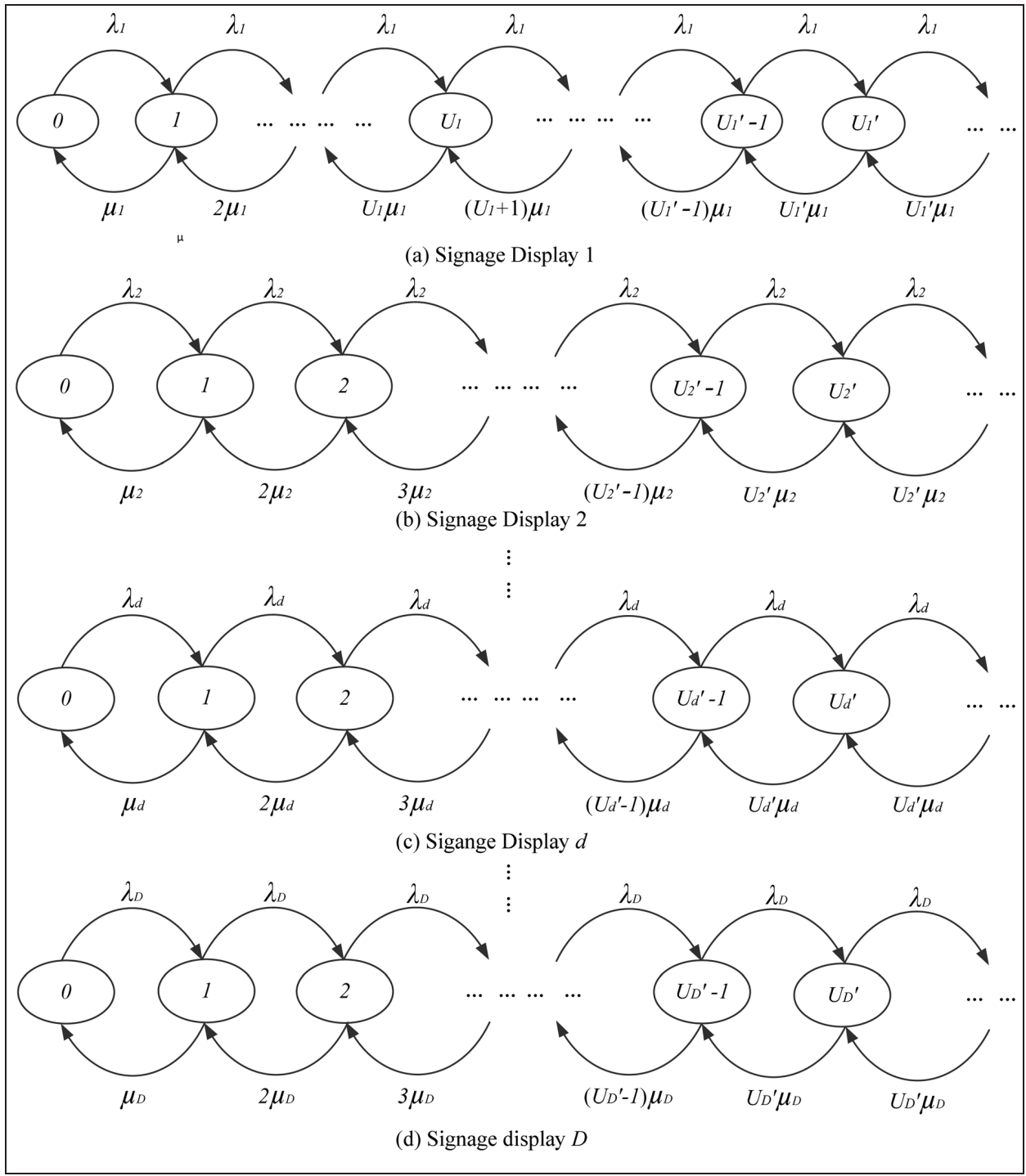

**Figure 6.** Markov chain diagram for requesting user after allocation.

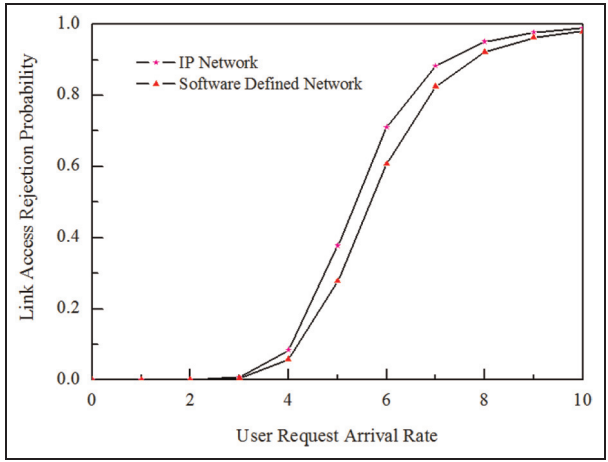

**Figure 7.** Link access rejection probability analysis.

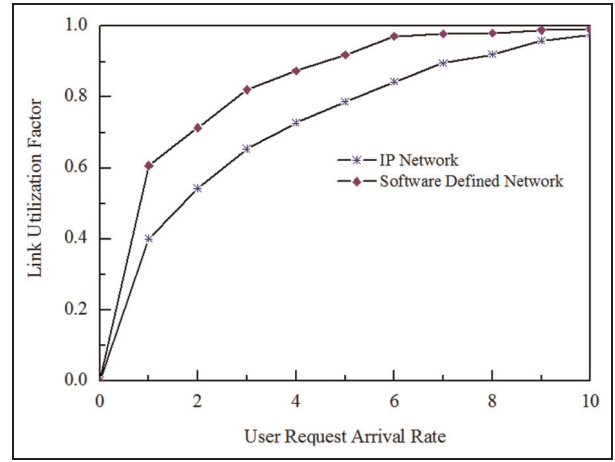

**Figure 8.** Comparison of link utilization factor.

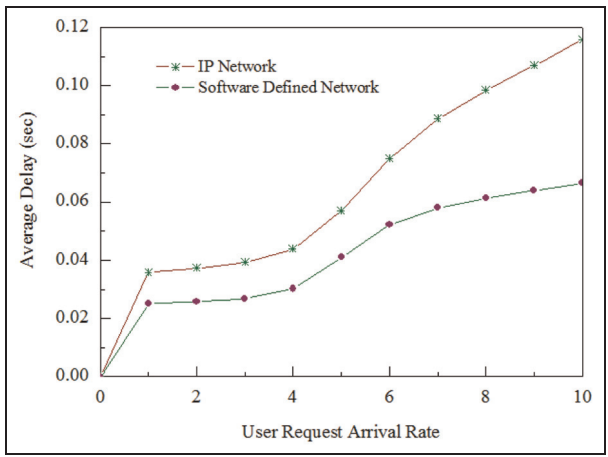

**Figure 9.** Delay profile comparison for SDN and IP network. SDN: software-defined networking; IP: Internet protocol.

**Table 2.** Parameters assumed for performance evaluation.

| Parameters | Value |
| --- | --- |
| Total number of signage display connected in a cloud server | 3 |
| Total reserved link channels in the cloud for each digital signage | 10 |
| Average duration of user in the link of the total system | 10 s |
| Request arrival rate | 1–10 per second |
| Ratio of the request arrival rate in the digital signage | 5:3:2 |

The delay profile of the proposed scheme has been shown in Figure 9, which indicates the average delay occurring for queued users in the cases of both the IP network and the SD network. The average delay is the average of the waiting time taken by all of the users of a network to connect to the links after sending requests. We have simulated the average delay based on the average time duration spent by every user in the link channels for consuming different services, because the users engaged in a link channel can cause a delay to another user to connect if there is no unused channel. However, the on-demand link allocation process of SD network can reduce the queue of the users waiting for accessing the web address. Therefore, the average delay time for the users waiting is reduced in the case of SD network while the IP network inherits higher average delay time. The increase in the number of the request arrival rate augments the average delay time as shown in Figure 9. The result signifies the advantage of the SD network over the IP network because the delay is crucial when the QoS of any system is measured. The rules update and the network status collection mechanism of the SD network help the system to reallocate the unused channels which reduce the delay time.

Another part of our performance analysis deals with the invisible ISC technology. However, the fundamental issue of our proposed invisible communication technology is the invisibility during the communication process, as it is imperative to keep the user unaware of the embedded data in the digital signage display. The visibility of the data in the display image can distort the image quality. With a view to analyzing the invisibility of our proposed scheme, a histogram analysis of the RGB colors of a prototype image has been performed that has been used in our implementation work. Figure 10 shows both the original and data-embedded images in (a) and (b), respectively, while the histogram of the RGB colors of those images has been represented in Figure 11. The $x$-axis of Figure 11 shows the intensity of the pixel within a range of 0–256, because an image of depth of 24 bit is used in which every color (RGB) has a depth of 8 bit. Although the images exhibited in the Figure 11 have no conspicuous dissimilarity to the eye of a user, the histogram illustrates a slight difference between the images with respect to the RGB colors. This minor difference in the intensity of the colors can be distinguished only by a camera. It is apparent from Figure 11 that the intensity of the red and blue colors has been changed very slightly, while the green color has remained unchanged. Consequently, the change in the intensity of colors has no effect on the quality of the image or what is seen by the human eye.

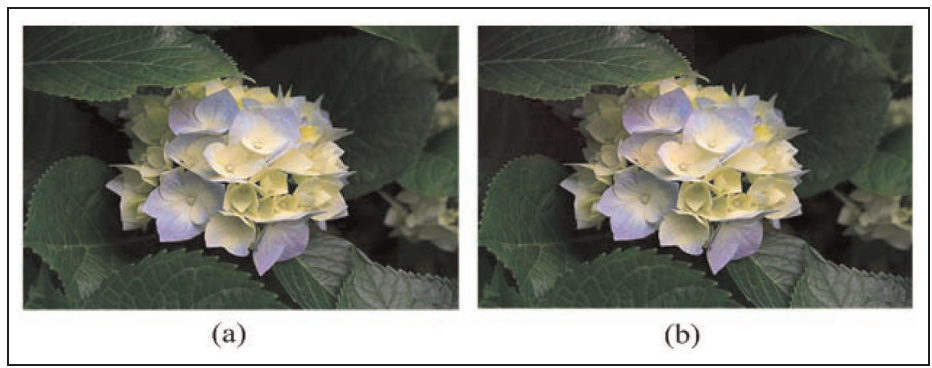

**Figure 10.** Image used in implementation and histogram analysis.

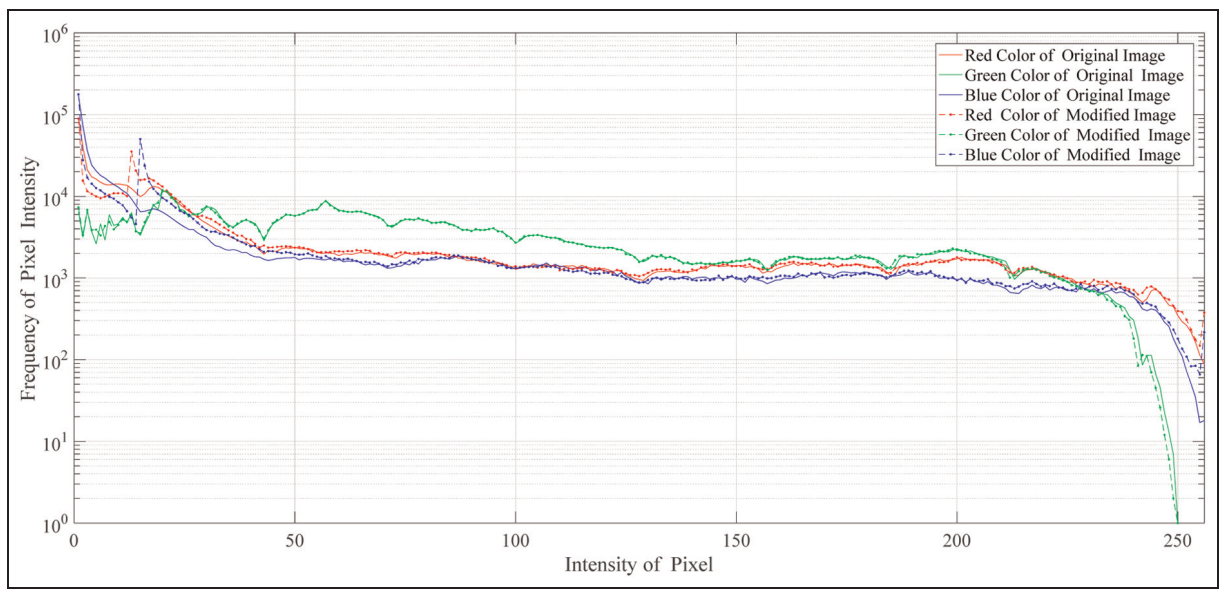

**Figure 11.** Histogram analysis of RGB part of the original image and the data-embedded image. RGB: red, green, and blue.

## Conclusion and future work

The maintenance of the increment in the number of users with limited resources on any network is a demand of time and is tough without any newer technologies. The concept of IoT has multifold the number of users in a network. The convergence of the SDN and invisible ISC technologies in our proposed scheme creates an opportunity to manage the increment in the number of users with the enhancement of QoS and to establish interactivity between the user and digital signage for initializing an IoT network. The proposed smart digital signage system is an illustration of SDN-based IoT architecture where the SDN controls the networking part, and the invisible ISC technology establishes the connection between the user and digital signage. The SD-IoT has the capability of managing the users with desired QoS, which expresses the compatibility of our proposed scheme in case of a huge number of users. The invisible ISC has retained the interactivity between the user and a display which can be a new way for device-to-human communication. Besides, the invisible ISC technology can be used as an energy-efficient method because there is no need to use the RF band in the communication. Besides, this communication can also be safe for human health. However, this smart system can be converged with another system to create an augmented interactivity. Our future works include the convergence of the smart digital signage system with other technology and the improvement of communication technology. The convergence can result in an effective IoT system with massive coverage in the connectivity of users in the system by inducing new bandwidth as well as can make the world smarter.

## Declaration of Conflicting Interests

## Funding

The author(s) disclosed receipt of the following financial support for the research, authorship, and/or publication of this article: This work was supported by the ICT R&D program of MSIP/IITP.(B0101-15-0429, Development of Open Screen Service Platform with Cooperative and Distributed Multiple Irregular Screens.)